\documentclass[%
preprint,
 amsmath,amssymb,
 aps,prl,
]{revtex4-2}

\usepackage{graphicx}
\usepackage{dcolumn}
\usepackage{bm}
\usepackage{xcolor}

\begin{document}

\preprint{APS/123-QED}

\title{\textcolor{black}{Nonlinear skin effect regime when a radio frequency electromagnetic field penetrates into a background plasma}}

\author{Haomin Sun$^{1}$, Jian Chen$^{2,*}$, Alexander Khrabrov$^{3}$, Igor D. Kaganovich$^{3}$, Wei Yang$^{4}$, Dmytro Sydorenko$^{5}$, Stephan Brunner$^{1}$}
\affiliation{$^1$Ecole Polytechnique F\'ed\'erale de Lausanne (EPFL), Swiss Plasma Center (SPC), CH-1015 Lausanne, Switzerland\\
$^2$Sino-French Institute of Nuclear Engineering and Technology, Sun Yat-sen University, Zhuhai 519082, People’s Republic of China\\
$^3$Princeton Plasma Physics Laboratory, Princeton University, Princeton, New Jersey 08543, USA\\
$^4$College of Physics, Donghua University, Shanghai 201620, People’s Republic of China\\
$^5$University of Alberta, Edmonton, Alberta T6G 2E1, Canada\\
$*$Corresponding author: chenjian5@mail.sysu.edu.cn
}


\vspace{10pt}


             
\begin{abstract}
  \textcolor{black}{Two-dimensional, electromagnetic particle-in-cell simulations are employed to study particle kinetics and power deposition in the skin layer when a Radio Frequency (RF) electromagnetic field penetrates into a background plasma. We identify a new regime at low frequency ($\sim\mathrm{MHz}$) and low pressure, where the motion of electrons can be highly nonlinear in the skin region. Through most of the RF cycle, the electrons are trapped in the effective potential formed by the vector and electrostatic potentials, with energy deposition being small and magnetic moment $\mu$ no longer being an adiabatic invariant. However, for a brief period around the null of the oscillating magnetic field, the electrons get detrapped, causing a jet-like current penetrating into the bulk plasma. During these brief periods, the power deposition becomes high, exhibiting a periodic burst nature.} Based on kinetic theory, we provide analytical expressions for the plasma current and energy deposition in the new regime. A criterion for transition between the newly identified low-frequency, periodic-burst regime and the usual anomalous non-local skin effect regime is proposed and verified. 

\end{abstract}

\maketitle


\section{}
\textbf{\textit{Introduction--}} \textcolor{black}{The nonlinear interaction between a time-varying Radio Frequency (RF) electromagnetic field and the background plasma has long been a topic of great interest, both in low temperature plasmas \cite{lieberman1994_book,Kolobov_95_nonlocal,Godyak_1999_skin_exp,Godyak_1999_PRL_secondharmonic,Smolyakov_2001_ponderomotive,Igor2002PRL_CCP,Kolobov_2017_lowfrequencyICP,Kolobov_19_kinetics} and space plasmas \cite{Yoon_whistler,TAO_JGR_whistler,Stenzel_06_Whistler,Akira_76_KAW_prl,Lin_2012_KAW_prl,Hoshino_2012_PRL,Hoshino_2015_PRL}. Studying the physics behind the skin effect in this interaction is crucial for understanding laboratory plasma generation systems \cite{lieberman1994_book,Chabert_Braithwaite_2011_ICPbook}, such as Capacitively Coupled Plasma (CCP) \cite{Turner95_CCPheating,Turner_2009review,kawamura2006_sto_CCP,Mussenbrock_2008_PRL_CCP,Liu_PRL_CCP,Zhao_2019_PRL_CCP,Mussenbrock_2008_PRL_CCP,Sun_2023_CCP_direct,JingyuSun_PRL2024,Yangyang_PRL2025} and Inductively Coupled Plasma (ICP) discharges \cite{Turner_93_ICPheating,Kolobov_95_EVDF_ICP,Kolobov_96_nonlocal_ICP,Kortshagen_95_ICP,Kaganovich_2003_kinetic_fullmodel_skin,Tuszewski_prl_real,Tyshetskiy_2003_PRL_Reductionheating_skin,Lee_APLreview,Lee_ICPEH_exp,Han3dmeasure2019,Han_3dexp,Gekelman2023}. The most classical skin effect in plasma refers to the condensation of electron current near the surface \cite{lieberman1994_book}. When an electromagnetic field penetrates into a background plasma, the classical skin depth is $\delta\sim c/\omega_{pe}$, where $c$ is the speed of light and $\omega_{pe}$ is the electron plasma frequency \cite{lieberman1994_book,Lieberman1998_CCPTOICP}. This theory assumes a local interaction between electromagnetic field and the plasma, describing the plasma skin behavior well at high gas pressure \cite{Subramonium_2004_3DICP,hsu2006_fluidICP,corr2008_fluidICP,bukowski1996_fluidICP}. Driven by experimental demands to achieve atomic-scale precision, plasma generation systems tend to operate at low gas pressure (few mTorr)  \cite{Chambers_52_anomalous,VIKolobov_1997_anomalousskin,Godyak_1998_PRL_expskin,Godyak_1999_skin_exp,Godyak_2000_skin_exp,Godyak2001_ponderomotive_exp,Liu_PRL_CCP,kawamura2006_sto_CCP,sharma2022_CCP,Sun_2023_CCP_direct,Sun_2022_PRE_beam,Sun_2022_PRL_beam,Xu_2023_spokes}, making it crucial to understand the underlying electron dynamics and the associated skin effects \cite{Polomarov_2006_selfconsistent_model_skin,Lee2018review}.} 

For low-pressure, high-frequency ($f=13.56~\mathrm{MHz}$ or higher) plasma generation systems, the skin effect was found to be of anomalous type \cite{VIKolobov_1997_anomalousskin,Godyak_1998_PRL_expskin,Godyak_1999_skin_exp,Godyak_2000_skin_exp,Godyak2001_EVDF_exp,Godyak2001_ponderomotive_exp,FFChen2001_nonlinearskin,Evans_2001_nonlocal_skin,Smolyakov_2000_nonlinear_skin,Tyshetskiy_2002_skin,Jiang_2009_hystersis,Lee2013_hystersis,El-Fayoumi_1998_lowfrequencyICP,Xu_2001_lowfrequencyICP,Kolobov_2017_lowfrequencyICP,Gao_POP_LowfrequencyICP,Isupov_2023_ferromagnetic,Xu_2001_lowfrequencyICP,Ashida_1996_pulsedICP,Cunge_1999_pulsedICP,Ramamurthi_2002_pulseICP,subramonium2002_ICPmodeling1,subramonium2002_ICPmodeling2,banna2012pulsed_ICP,Walter2023APSDPP,fantz2006spectroscopy_ICPfornegativeion,speth2006overview_ICPfornegativeion,Li_2019_negative_ion,Zielke_2022_lowfrequencyICP}, meaning that the skin depth,  $\delta$, is determined by non-local kinetics as $\delta\sim (v_{th}c^2/\omega\omega^2_{pe})^{1/3}$ instead of $\delta\sim c/\omega_{pe}$, where $v_{th}$ is electron thermal velocity, $\omega$ is the driving frequency. Electron heating in the anomalous regime is mainly due to wave-particle resonance of low-energy electrons, which can be treated similarly to the Landau damping \cite{Kaganovich_2003_kinetic_fullmodel_skin,Kaganovich_2004_anisotropic_skin,Kaganovich_2004_Landaudamping_skin,Tyshetskiy_2002_skin,Smolyakov_2003_nonlinear_skin,Tyshetskiy_2003_PRL_Reductionheating_skin,Polomarov_2006_selfconsistent_model_skin,Lee_2012_experiment_skin}. The anomalous regime was also called the non-local regime because over one RF period the electrons travel a significant distance $d\sim v_{th}/\sqrt{\omega^2+\nu^2}>\delta$  \cite{Tyshetskiy_2002_skin}, where $\nu$ is the collision frequency. As a result, the plasma current is a non-local function of the RF electric field.
Multiple modeling studies have been performed for the plasma generation systems in this regime, utilizing the non-local kinetic approach \cite{Kaganovich_2003_kinetic_fullmodel_skin,Badri_Ramamurthi_2003_ICP,Polomarov_2006_selfconsistent_model_skin,froese_thesis_2007_1DPIC_skin,ICPYang_2021_PPCF,Yang_2022_Chambersizeeffect,YANG_2022_Conductivityeffect,Yang_2022_Excitedstateeffect} 
to predict the spatial distribution of the plasma current and power deposition. 
\textcolor{black}{At lower frequency ($\sim 1$MHz), however, it is shown in experiments that the skin effect in plasma generation systems manifests behavior of nonlinear type \cite{Godyak_1999_PRL_secondharmonic} rather than anomalous type, which is featured by the second harmonics in electron current and electrostatic potential \cite{Godyak_1999_PRL_secondharmonic,Godyak_2000_skin_exp,Godyak_1999_skin_exp,Kolobov_19_kinetics}. }

\textcolor{black}{However, due to the lack of self-consistent kinetic 2D modeling, the physics of the nonlinear skin effect at low frequency \cite{Godyak_1999_PRL_secondharmonic,Godyak_1999_skin_exp} is not sufficiently understood}, because in experiments it is difficult to perform detailed measurements of plasma properties and electromagnetic fields within the skin layer and correspondingly deduce particle kinetics  \cite{Kolobov_2017_lowfrequencyICP,Godyak2001_EVDF_exp}. Although it was previously shown that at low frequencies the particle dynamics could become highly nonlinear  \cite{Cohen96_RF_Bfieldeffects1,Cohen96_RF_Bfieldeffects2}, those studies could not be applied here because the RF magnetic field and electrostatic potential need to be determined self-consistently. Existing kinetic simulations did not focus on the nonlinear skin effect at low RF frequency \cite{GIBBONS_1995_DarwinPIC_ICP,Turner_1996_ICP1DPIC,Sydorenko_2005_PIC1d,froese_thesis_2007_1DPIC_skin,Fu_2024_2DPIC,Wen_2025}.


\textcolor{black}{We demonstrate for the first time that a strongly nonlinear skin effect regime with periodic bursty energy deposition presents at low RF frequency. The electrons are trapped in effective potential, with magnetic moment $\mu$ no longer being an adiabatic invariant.
 Detrapping occurs when the RF magnetic field passes through zero, and electrons quickly escape the area near the coil to form a current jet, causing periodic bursts in the energy deposition. This nonlinear skin effect causes a higher third harmonics in the induced electron current. A kinetic theory is developed and predicts well the plasma current in this bursty regime. To our knowledge, this work presents the first comprehensive 2D PIC simulations for the nonlinear skin effect in the interaction between plasma and RF electromagnetic field. }

\begin{figure}
    \centering
    \includegraphics[width=0.87\textwidth]{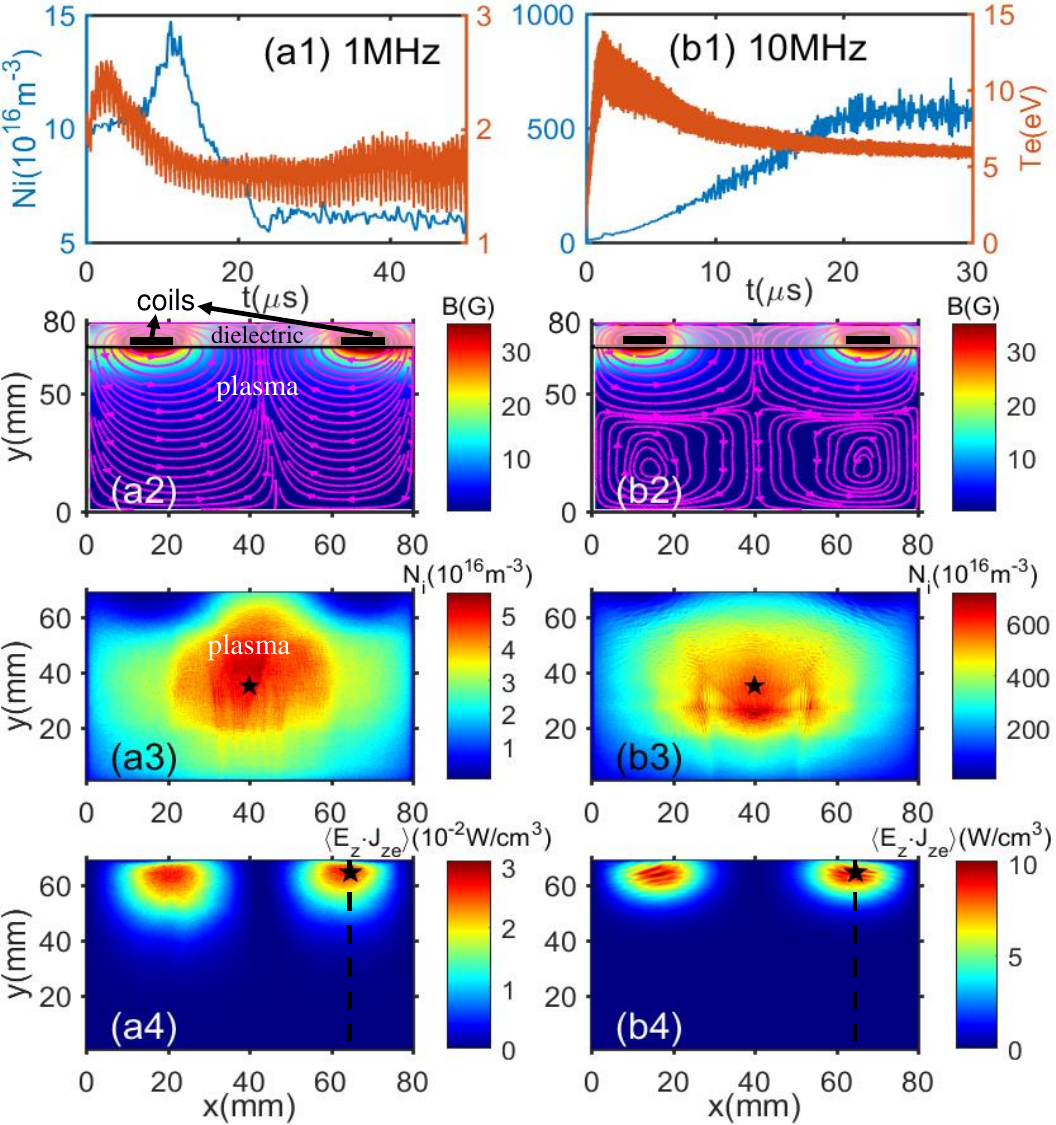}
    \caption{Time evolution of physical quantities and steady-state profiles. Panel (a) is for the low frequency case $f=1\mathrm{MHz}$ and panel (b) is for $f=10~\mathrm{MHz}$ case. Subplots (a1) and (b1) show time evolution of the ion density and electron temperature. The probes are placed at the center, shown by black stars in (a3) and (b3). Subplots (a2) and (b2) show 2D maps of the magnetic field strength with superimposed magenta curves tracing the field lines. Black rectangles are the cross-sections of the coil wires, and the black horizontal line is the plasma facing boundary of the dielectric. Subplots (a2) and (b2) are plotted at the phase when the RF magnetic field is at the maximum. Subplots (a3)-(b4) are the time-averaged (denoted by $\langle...\rangle$) ion density and energy deposition.}
    \label{fig1_timetracesnapshot}
\end{figure}

\textbf{\textit{Method --}} 
\textcolor{black}{We take ICP discharge system as an example to study the nonlinear skin effects.} We implemented the electromagnetic Darwin scheme to EDIPIC-2D \cite{sydorenko2024improvedalgorithmtwodimensionaldarwin,Charoy_EDIPIC_benchmark,Sun_2022_PRE_beam,Sun_2022_PRL_beam,Jin_2022_psst,Cao2023pop,Chen2024_intermittencypop,Jin_2024_psst}. The simulations are performed in Cartesian geometry. Figure \ref{fig1_timetracesnapshot} (a2) and (b2) depict the simulation domain. The rectangular chamber dimensions are $D_x\times D_y=80~\mathrm{mm}\times 80~\mathrm{mm}$, with a dielectric slab located at $70~\mathrm{mm}<y<80~\mathrm{mm}$ and antennas positioned within it (black rectangles showing cross-section). The plasma occupies the rest of the simulation domain. Because the code implements a direct-implicit time advance \cite{COHEN_1982_DI,GIBBONS_1995_DarwinPIC_ICP,Sun_2023_CCP_direct,sydorenko2024improvedalgorithmtwodimensionaldarwin}, we set the cell size to be $\Delta x=0.33~\mathrm{mm}$, resulting in overall grid dimensions of $N_x\times N_y=240\times 240$. The antenna currents are $180^{\circ}$ out of phase, oscillating with the frequency $f=\omega/2\pi$ in the range of $1$ to $10~\mathrm{MHz}$ and the amplitude $I_{coil}=60~\mathrm{to}~280~\mathrm{A}$. \textcolor{black}{The coils create an RF electromagnetic field that penetrates into the background plasma.} The amplitude of the coil current is relatively high because it is necessary to sustain the plasma. Similar values have been used in experiments \cite{Godyak_1999_skin_exp,Godyak_2000_skin_exp,Mattei_2016_ICPEH_exp}. The domain is surrounded by conductive boundaries with secondary emission neglected. Initially, the system contains Argon gas at the pressure $p=5~\mathrm{mTorr}$ and a plasma with a uniform density $N_e=N_i=10^{17~}\mathrm{m}^{-3}$, electron temperature $T_e=2.0~\mathrm{eV}$, and ion temperature $T_i=0.03~\mathrm{eV}$. $1000$ macro-particles per cell are specified initially for all species. Each simulation is analyzed after reaching a steady state.

\textbf{\textit{Results--}}We first focus on two cases, with $I_{coil}$ at $130A$ and the frequencies of $f=1~\mathrm{MHz}$ and $f=10~\mathrm{MHz}$. Figure~\ref{fig1_timetracesnapshot} shows the time histories and spatial profiles of the plasma for these cases. The time evolution of the ion density and electron temperature shown in Fig.~\ref{fig1_timetracesnapshot} (a1) and (b1) indicates that the simulations have reached a steady state. Subplots (a2) and (b2) show the magnetic field lines with superimposed color-maps of the field strength. The magnetic field amplitude is almost the same between the low- and the high-frequency cases because it depends mainly on the coil current ($B\propto I_{coil}$). 
Comparing the ion density profiles between Fig.~\ref{fig1_timetracesnapshot} (a3) and (b3), we see a significant difference. This is due to a comparable difference in energy deposition between (a4) and (b4), because $E_z\cdot J_{ze}\propto \omega^2I_{coil}$. 

\begin{figure}
    \centering
    \includegraphics[width=0.87\textwidth]{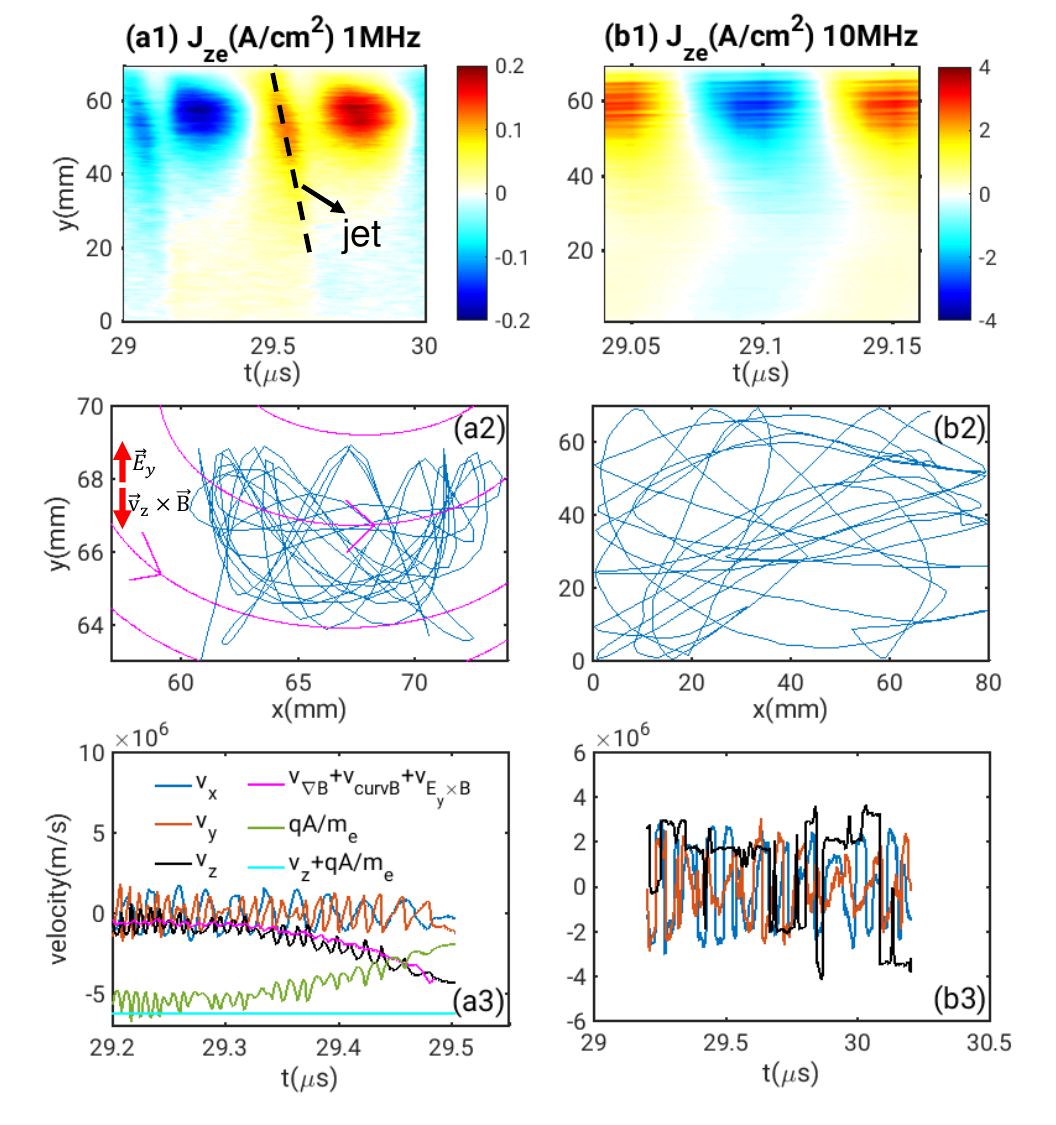}
    \caption{Subplots (a1) and (b1) show the color maps of the current density vs $y$ and $t$; the domain cross sections are taken along the black dashed lines in Fig.~\ref{fig1_timetracesnapshot} (a4) and (b4), respectively. Subplots (a2) and (b2) show representative particle trajectories, respectively, for the $1~\mathrm{MHz}$ and $10~\mathrm{MHz}$ cases over the time intervals marked in subplots (a3) and (b3). The red arrows in (a2) designate the counteracting electrostatic and Lorentz forces acting upon the particle in the $1~\mathrm{MHz}$ case. Subplots (a3) and (b3) show the time evolution of the particle velocity and the drift velocity components, with $v_{\nabla B}$ being the drift due to the magnetic field gradient, $v_{curvB}$ being curvature drift, and $v_{E_y\times B}$ being $E_y\times B$ drift. The vector potential, expressed as $qA/m_e$, and the canonical momentum, $v_z+qA/m_e$ for $1~\mathrm{MHz}$ case are traced by the green and light blue lines.}
    \label{fig2_particletrace}
\end{figure}

A close examination of the two cases shown in Fig.~\ref{fig2_particletrace} reveals two notable distinctions. First, a pronounced jet-like structure of the electron current is observed in the time history of the low-frequency case (see Fig.~\ref{fig2_particletrace} (a1)). The jet-like structure propagates into the plasma at around $t=29.5\mu s$, with a speed close to the electron thermal velocity $v_{th}$. \textcolor{black}{The penetration of electron current into plasma in Fig. \ref{fig2_particletrace} (b1), however, is due to anomalous skin effect and shows a clear qualitative difference from Fig. \ref{fig2_particletrace} (a1).} The second distinction is in the representative particle trajectories shown in Fig.~\ref{fig2_particletrace} (a2) and (b2) for low- and high-frequency cases. \textcolor{black}{For low-frequency case, electrons are 
trapped in effective potential (see End Matter)} with a bounce-radius ($\sim 2\mathrm{mm}$) smaller than the skin depth ($\sim 10~\mathrm{mm}$). The electrons are, however, not trapped for the high-frequency case. 
To track the electrons, we randomly sampled 400 particle trajectories originating from $x=65.0~\mathrm{mm}$ and $y=65.7~\mathrm{mm}$ (the location marked by the black stars in Fig.~\ref{fig1_timetracesnapshot} (a4) and (b4)) at $t=29.2~\mu s$. We choose this location and time because in the $1~\mathrm{MHz}$ case, most new electrons are produced by ionization near the coil at this phase of the RF period (see Fig.~3 in our accompanied paper \cite{Sun_ICP_2025}). 
A movie for this particle is available in the supplemental material \cite{supplemental}. 
This particle behavior is in significant contrast to the findings of Ref.~\cite{Froese_POP_Nonlinearskin}, where the Larmor radius was assumed to be much larger than the skin depth. It is evident from Fig.~\ref{fig2_particletrace} (a2) that electrons are \textcolor{black}{trapped near the coil region}. The oscillations of velocities shown in (a3) indicate that the electron undergoes a cyclotron motion in $y-z$ plane and bounces parallel to the magnetic field in $x$~direction. 
It is important to note that the electrons' $y$ locations remain almost unchanged. This is due to the plasma-generated ambipolar electric field \cite{Dima2005POPponderomotive}, $E_y$, 
counterbalancing the Lorentz force $\Vec{v}_z\times \Vec{B}$ and making the time-averaged net force on the electrons in $y$ direction vanish. \textcolor{black}{Therefore, the bounce invariant $I=\int v_y dy$ instead of $\mu=m_ev_{\perp}^2/2B$ is an adiabatic invariant for the trapped electrons (see End Matter).} Figure~\ref{fig2_particletrace} (a3) further shows that the particle motion in $z$~direction is due to particle drifts, with $E_y\times B$ drift being dominant \cite{Sun_ICP_2025}.
As the RF magnetic field is $\pi/2$ out of phase with the inductive electric field and it decreases as the electric field increases, the electrons eventually \textcolor{black}{detrap from effective potential near the phase when the magnetic field becomes small}: the electrons move into the plasma interior at around $t=29.5~\mu\text{s}$, forming a jet-like structure shown in Fig.~\ref{fig2_particletrace} (a1). During this process, the electron's canonical momentum $m_ev_z+qA$ is conserved (denoted by the light blue line in Fig.~\ref{fig2_particletrace} (a3)), where $A$ is the vector potential. A sample test particle in the $f=10~\mathrm{MHz}$ case, however, travels over the entire simulation domain, interacting with the coil field in a transient way, as described in multiple previous works \cite{Cohen96_RF_Bfieldeffects1,Cohen96_RF_Bfieldeffects2}.

\textbf{\textit{Theory--}}As shown in Fig. \ref{fig2_particletrace} (a3) and justified in the End Matter, electron motion can be treated as a cyclotron rotation in $y-z$ plane around a guiding center drifting in $z$ direction. 
To obtain an analytical expression for electron current, the electron distribution function $f$ is split into a background Maxwellian $f_0$ as function of local electron density and temperature; and $f_1$, a perturbation to it. We use the assumption of: 1) The maximum electron bounce frequency (in effective potential) $\Omega_{max}\gg\omega$. Indeed, for $1~\mathrm{MHz}$ case $\Omega_{max}\sim 10^{8}~\mathrm{s}^{-1}$ and $\omega=6.28\times 10^6~\mathrm{s}^{-1}$. 
2) We only consider a 1D spatial variation, specifically along the black dashed line in Fig.~\ref{fig1_timetracesnapshot} (a4). Similar assumptions have been made in the pioneering paper of Tuszewski \cite{Tuszewski_prl_real}. \textcolor{black}{The theory considers RF magnetic field and electrostatic potential, and is the first to predict current that can be compared with PIC simulations when nonlinear skin effect dominates.}
 
The kinetic equation for the perturbed electron distribution function $f_1$ is 
\begin{equation}\label{unperturbed1}
    \frac{\partial f_1}{\partial t}+\Vec{v}\cdot\frac{\partial f_1}{\partial\Vec{r}}+\frac{q}{m}(\Vec{v}\times\Vec{B}+\Vec{E}_{sc})\cdot\frac{\partial f_1}{\partial\Vec{v}}=\frac{q}{m}\frac{\partial\Vec{A}}{\partial t}\cdot\frac{\partial f_0}{\partial\Vec{v}},
\end{equation}
where $\Vec{A}$ is the vector potential of the EM field and $E_{sc}$ is the longitudinal electric field $E_y$ resulting from charge separation. Note that we have $\vec{E}_{sc}\gg\partial\vec{A}/\partial t$ based on simulations. Essentially, we consider the case where the RF magnetic field strongly alters the particle trajectory. This is the opposite case to kinetic theories describing the high frequency case \cite{ICPYang_2021_PPCF}. Electron-neutral collision is neglected because the collision is weak $\nu_{en}=3.2\times 10^6~\mathrm{s}^{-1}<\omega\ll \Omega_{max}$ and does not affect our results significantly. 
The full derivation is presented in our accompanied paper \cite{Sun_ICP_2025}. The total current in $z$ direction is \cite{Qin_2000_gyroequilibrium,Lee2003_gyrokinetic}
\begin{equation}\label{totalcurrentreal}
    \vec{J}_{ze}=\int d\vec{v}(\vec{v}_{z,d}+\vec{v}_{\perp})f_0+\int d\vec{v}(\vec{v}_{z,d}+\vec{v}_{\perp})f_1=\vec{J}_{ze,0}+\vec{J}_{ze,1},
\end{equation}
where $\vec{v}_{z,d}$ is the drift velocity in $z$ direction and $\vec{v}_{\perp}$ is the perpendicular gyro-velocity. Upon integrating Eq.~\eqref{unperturbed1} along the unperturbed electron trajectory, we obtain an analytical expression for $\hat{f}_1$ (the Fourier transform of $f_1$). The approximation to the electron current $J_{ze,1}$ obtained by integrating $\hat{f}_1$ is

\begin{multline}\label{Eq_currentreal}
    J_{ze,1}(t)=Re\left\{\sum_{k_y}\hat{J}_{ze,1}(t)e^{ik_y y}\right\}=Re\left\{\sum_{k_y}\left[\frac{\omega q^2N_e\hat{A}(t)}{T_e}v'_d\frac{v_F}{\omega}e^{-\xi}I_0(\xi)- \right.\right.  \\
    \left.\left.   i\frac{\omega q^2N_e\hat{A}(t)}{T_e}\frac{(v'_d+v_F) k_y v^2_{th}}{\Omega\omega}e^{-\xi}\left(I_0(\xi)-I'_0(\xi)\right)-\right.\right.
    \left.\left.\frac{\omega q^2N_e\hat{A}(t)}{T_e}\frac{v^2_{th}}{\omega}e^{-\xi}\left(2\xi I_0(\xi)-2\xi I'_0(\xi)\right) \right]e^{ik_y y}\right\},
\end{multline}
where $Re\left\{...\right\}$ denotes the real part, $\hat{J}_{ze,1}$ and $\hat{A}$ denote the Fourier components, $N_e$ and $T_e$ are the time-averaged local electron density and temperature, both from definition of $f_0$. Also, $\Omega(t)\approx qB(t)/m_e$, $v_F\approx -\langle F\rangle/qB(t)+E_y/B(t)$ is the total electron drift velocity (where $E\times B$ drift is dominant), ${\langle F\rangle}=-\langle m_e(v^2_{\perp}/2+v^2_{||})\rangle\nabla_{\perp}lnB$ is the average force (over particles) due to the magnetic field gradient and curvature, $v_{\perp}$ and $v_{||}$ are the gyro-velocity components in cylindrical coordinates, $I_0$ is the $0\mathrm{'s}$ order modified Bessel function of the first kind (the prime on it denotes derivative), $\xi=(k_yv_{th}/\Omega(t))^2$, where $k_y$ is the argument of the Fourier component of ${A}$, and the diamagnetic drift velocity is $v'_d\approx T_e/qB(t) dlnN_e/dy$. The first term in Eq.~\eqref{Eq_currentreal} is due to the drift $\vec{v}_{z,d}$, whereas the second and third terms represent contributions from $\vec{v}_{\perp}$. The guiding center current $J_{ze,0}$ related to $f_0$ comes mainly from $E_y\times B$ drift
\begin{equation}\label{Eq_f0contri}
    J_{ze,0}(t)=qN_ev'_F\left(1-\exp\left(-\frac{1}{2}\frac{v^2_{\perp,tr}}{v^2_{th}}\right)\right),
\end{equation}
where $v_{\perp,tr}=|qB(t)|\delta/2m_e$ is the maximum perpendicular velocity for the electrons to be trapped by the magnetic field, since only the trapped electrons in $f_0$ contribute to the current. And $v'_F=E_y/B(t)$. In Eq.~\eqref{Eq_f0contri} we have neglected the diamagnetic current since it is much smaller than the $E\times B$ current.

Our case is more complicated than the standard gyrokinetic treatment \cite{Lee2003_gyrokinetic,Caryhamitoniantheory,Stephanlecturenote} in two aspects. First, only the electrons are magnetized, so the current from electron $E_y\times B$ drift dominates and cannot be canceled with that of the ions. Second, the guiding magnetic field is time-varying and the drift velocity $v_F$ is comparable to or larger than the thermal velocity $v_{th}$, making the terms associated with $v_{F}$ dominant. Because of this, the contribution from $f_1$ to the current is not negligible.
Equation~\eqref{Eq_currentreal} for $J_{ze,1}$ contains Bessel functions. Those represent the Finite Larmor Radius (FLR) effect.
Physically, the FLR effect occurs because when a particle undergoes gyro-motion and a field gradient is present, the particle tends to spend more time on the low-field side than the high-field side \cite{Littlejohn_1983,CARY1983,Hahm_gyrokinetic_7246860,Caryhamitoniantheory}. Effectively, the field acting on the electrons will be reduced, hence reducing the current. Therefore, $J_{ze,1}$ takes the sign opposite to that of $J_{ze,0}$ and acts to reduce the current. 
The dominance of the RF magnetic field on particle motion significantly changes the electron current (see simplified formula Eq. (12) in Ref. \cite{Sun_ICP_2025}), indicating the presence of a new physical regime.
With the gradual increase of coil current, the plasma density increases and the electron trajectory changes to the classical type as seen in Fig.~\ref{fig2_particletrace} (b3), with all drifts becoming negligible and the electron current taking the form in existing kinetic theory $\hat{J}_{ze,old}=\omega q^2/m_e\int v_z/(\omega-k_yv_y)\partial f_0/\partial \vec{v}\cdot\hat{A}d\vec{v}$ \cite{Kaganovich_2003_kinetic_fullmodel_skin,ICPYang_2021_PPCF}. 
\begin{figure}
    \centering
    \includegraphics[width=0.85\textwidth]{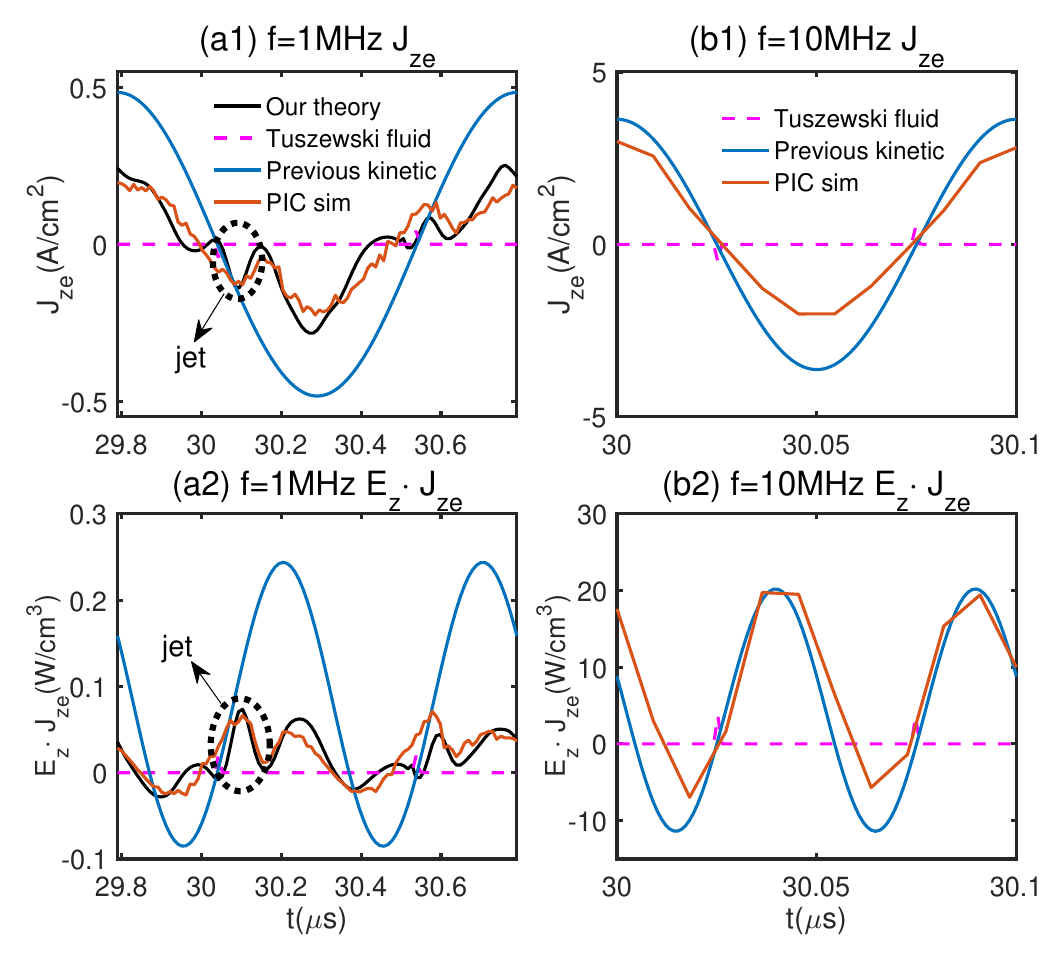}
    \caption{Comparing PIC simulations with theory for (a): $1~\mathrm{MHz}$ case and (b): $10~\mathrm{MHz}$ case. Subplots (a1) and (b1) show the electron current $J_{ze}$ and subplots (a2) and (b2) show the energy deposition rate. The comparison is made at the locations denoted by black stars in Fig.~\ref{fig1_timetracesnapshot} (a4) and (b4). For our kinetic theory, we use Eq.~\eqref{totalcurrentreal} with $k_y=2\pi n_y/L_y$ with $n_y$ from $1$ to $210$ (number of cells occupied by plasma) to compute $J_{ze,1}$. The fluid approach is from Ref.~\cite{Tuszewski_prl_real} and the previous non-local kinetic calculation is from Ref.~\cite{ICPYang_2021_PPCF}.}
    \label{fig3_theorycompare}
\end{figure}
Figure~\ref{fig3_theorycompare} compares the time evolution of $J_{ze}$ and energy deposition between the theoretical predictions and the PIC simulations. For our theory (black) and the Tuszewski theory (magenta dashed, fluid model) \cite{Tuszewski_prl_real}, we evaluate the electric field and other relevant quantities at the location marked by the black stars in Fig.~\ref{fig1_timetracesnapshot} (a4) and (b4). We then calculate the electron current from Eq.~\eqref{totalcurrentreal} and the conductivity given in Ref.~\cite{Tuszewski_prl_real}, respectively. 
To carry out a calculation based on the previous non-local kinetic theory developed for high frequency ICP (blue), we employ the code provided in Ref.~\cite{ICPYang_2021_PPCF} to perform a 2D simulation of the entire system, and then extract the electric field and electron current at the same location marked in Fig.~\ref{fig1_timetracesnapshot} (a4), (b4). This existing kinetic theory assumes uniform plasma density. We see that the fluid theory does not predict the current and the energy deposition rate, whereas the previous kinetic theory performs well only for $f=10~\mathrm{MHz}$ case. \textcolor{black}{At $1~\mathrm{MHz}$, our theory displays a much better agreement for both the current and energy deposition, \textit{while the other theories predict energy deposition that deviates from PIC simulations by at least a factor of $5$}.} \textcolor{black}{The electron current for $f=1\text{MHz}$ shows a higher third harmonic component (Fig. 15 in \cite{Sun_ICP_2025}). Physically, this is due to the coupling between second harmonic electron motion in $y$ direction (due to potential, see Fig. 37 in \cite{Sun_ICP_2025}) and RF magnetic field at fundamental frequency.}
We identify the new regime dominated by periodic bursty energy deposition as the ``periodic burst regime''. 
When averaging over the simulation domain, the jet-like current 
contributes around $70\%$ to the total energy deposition. 

\begin{figure}
    \centering
    \includegraphics[width=0.99\textwidth]{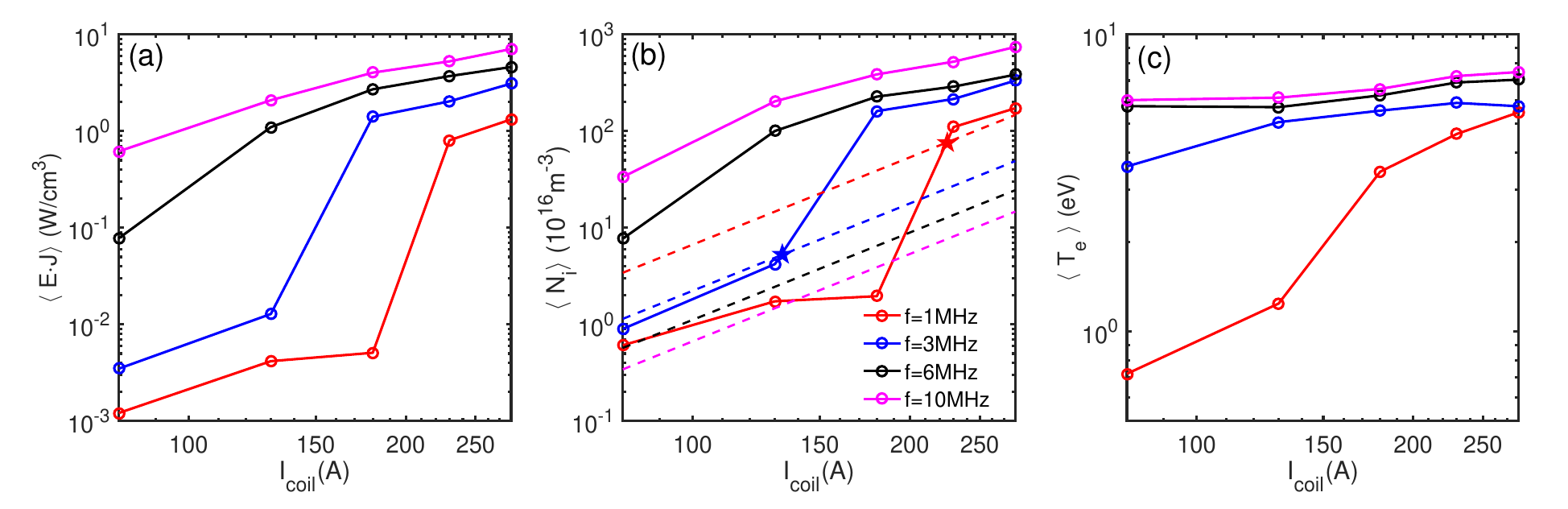}
    \caption{The solid lines show, in a steady state, (a) energy deposition rates, averaged over the entire simulation domain and RF period, (b) time-averaged ion density at the probe location denoted by the black star in Fig.~\ref{fig1_timetracesnapshot} (a4), and (c) time-averaged electron temperature at the probe location denoted by the black star in Fig.~\ref{fig1_timetracesnapshot} (a3), for different values of the coil current and driving frequency. The dashed lines denote the onset threshold for the periodic burst regime, which follows the scaling law of $N_e\sim I_{coil}^3/\omega T_e$ according to Eq.~\eqref{estimte}. Each dashed line denotes the boundary for the solid line of the same color in (b). The relevant magnetic field and temperature data for calculating the dashed lines is evaluated at the black stars in Fig.~\ref{fig1_timetracesnapshot} (a4) and (b4). The red and blue stars in (b) denote the transition points between the two regimes.}
    \label{fig4_largescansummary2}
\end{figure}

\textbf{\textit{Parameter Scan--}}We perform a parameter scan over $I_{coil}$ and $\omega$ to identify the onset boundary for periodic burst regime. Figure~\ref{fig4_largescansummary2} shows the average energy deposition, ion density, and electron temperature for different values of the driving waveform parameters. 
We see that at lower frequencies there is a distinct jump in energy deposition. A similar jump is observed for the ion density. This transition occurs because in the periodic burst regime, the electric field and current are nearly out of phase during most of the RF period, since the strong RF magnetic field makes plasma response nearly local. Only in the phase of electron jet will the energy deposition become significant. Therefore, the energy deposition becomes lower than in the anomalous skin effect regime, making the electron density also much lower. Based on this understanding, the criterion for transition between the bursty regime and the anomalous non-local regime is estimated by (see also End Matter)
\begin{equation}\label{estimte}
    2\frac{m_ev_{th}}{|qB|}=\left(\frac{v_{th}c^2}{\omega \omega^2_{pe}}\right)^{1/3}.
\end{equation}

When the diameter of the electron gyro-circle is larger than the skin depth, the electrons are unlikely to undergo a full gyro-cycle, and are therefore no longer being confined by the effective potential. 
Based on Eq.~\eqref{estimte}, an estimate of the plasma density at the threshold of the periodic burst regime is obtained. 
The cases with different driving frequencies will have different transition values for the plasma density, as indicated in Fig.~\ref{fig4_largescansummary2} (b) by the dashed lines. 
One expects to obtain the periodic burst regime in the region located below the dashed lines. A reasonable agreement is found, indicating the validity of Eq.~\eqref{estimte} for finding the boundary between the two regimes.

\textbf{\textit{Conclusions--}}\textcolor{black}{We identify a new nonlinear skin effect regime when an RF electromagnetic field penetrates into a background plasma, where the energy deposition property manifests a periodic burst nature. In this regime, the electrons are trapped in effective potential, 
with $\mu$ no longer being the adiabatic invariant. When the RF magnetic field becomes weak, the electrons detrap, forming a jet-like current propagating quickly into the plasma and accounting for a sizable fraction of the energy deposition.   
As a result, the electron current in this regime shows higher harmonics instead of just a sinusoidal shape with basic frequency.
A kinetic theory is proposed to estimate the electron current, showing good agreement with PIC simulations.} The condition for the transition from anomalous skin effect regime (untrapped electrons) to the periodic burst regime is identified. 
A significant jump in the plasma density is observed during such transition, similar to what was observed in the E-H mode transition \cite{UKortshagen_1996_ICPEH_theory,Chabert_2003_ICPEH_theory,Lee_ICPEH_exp,Mattei_2016_ICPEH_exp,Wegner_2017_ICPEH_exp}. 
We propose an estimate for current density (Eq.~\eqref{totalcurrentreal}) and a criterion for transition to the periodic burst regime  (Eq.~\eqref{estimte}) that can be verified in future experiments.

\textbf{\textit{Acknowledgement--}} The authors thank Dr. Edward Startsev for checking the derivations in this paper. The authors thank Dr. Sarvesh Sharma for the discussions of high frequency harmonics. The work was partially supported by the National Natural Science Foundation of China (Grant No. 12305223) and the National Natural Science Foundation of Guangdong Province (Grant No.2023A1515010762). This research was also supported in part by the U.S. Department of Energy, Office of Fusion Energy Science under Contract No. DE-AC02-09CH11466. 

\appendix
\section{\textcolor{black}{End Matter: Electron motion in a 1D model of low-frequency ICP skin layer with account for strong RF magnetic field and electrostatic potential}}

\begin{figure}
    \centering
    \includegraphics[width=1.00\textwidth]{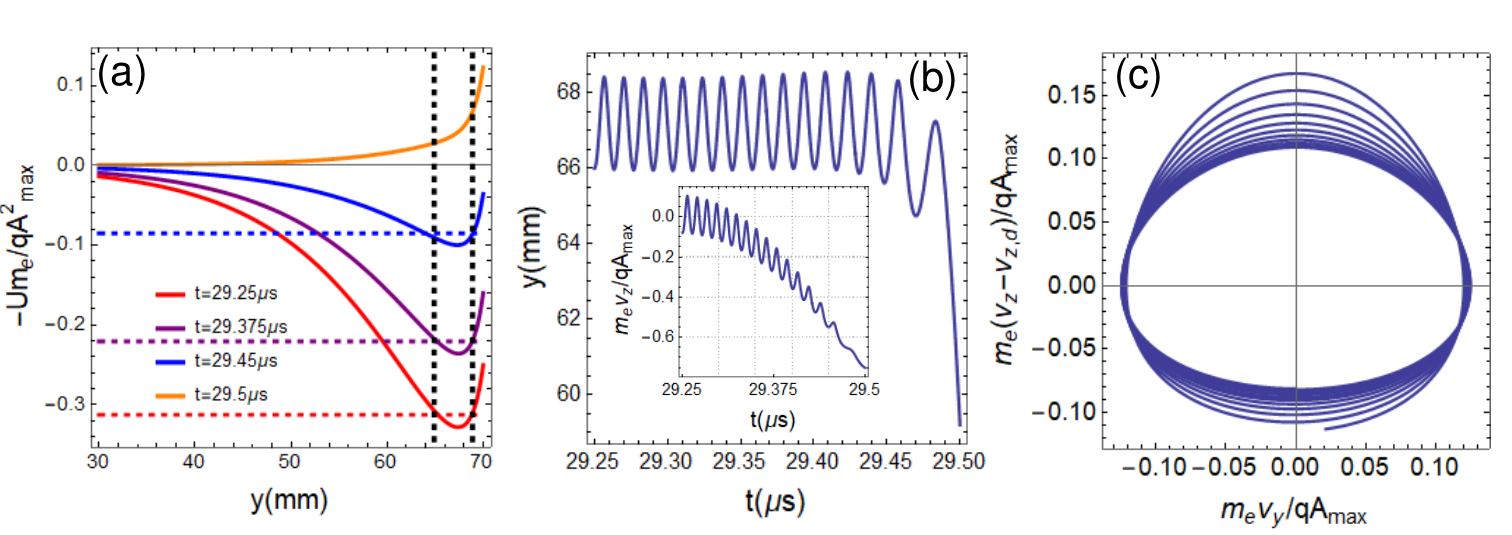}
    \caption{\textcolor{black}{Subplot (a) shows the time evolution of the normalized effective potential $U$, with the horizontal dashed lines denoting the electron energy level and the vertical dashed lines locating the turning points of an orbit trapped in the effective potential well. Note at $t=29.5\mu s$ the electron becomes un-trapped and escapes the skin layer towards the bulk plasma. Subplots (b) and (c) represent the trajectory of an electron with initial velocity $v_{y0}=-0.1qA_{max}/m_e$, $v_{z0}=0$, and initial position $y_0=66mm$. The velocity is normalized by $eA_{max}/m_e=7\times 10^6~\text{m/s}$.}}
    \label{fig_effectivepotential}
\end{figure}

\textcolor{black}{In this End Matter, we show that the electron trajectory in the low frequency $1\text{MHz}$ case is not simply affected by the RF magnetic field, but by the \textit{effective potential} formed by the combination of RF magnetic field and electrostatic potential. To better understand the representative electron trajectory shown in Fig.~\ref{fig2_particletrace} (a), we turn to a simplified 1D model of electron motion along the vertical dashed line in Fig.~\ref{fig1_timetracesnapshot} (a4). Note that the electron motion parallel to magnetic field is not considered because the bounce parallel to the magnetic field occurs with lower frequency, and is not expected to significantly affect the electron current. Here, the simulation data shows that the vector potential can be approximated as $A_z(y,t)=A_{max}\exp(-y/\delta)cos(\omega t)$ (where $\delta$ is the skin depth). The RF magnetic field in the $x$ direction is $B_x(y,t)=\partial A_z/\partial y$ and the RF inductive electric field in the $z$ direction is $E_z(y,t)=-\partial A_z/\partial t$. The electron equation of motion reads
\begin{equation}
    m_e\frac{dv_y}{dt}=qv_zB_x-q\frac{\partial \Phi}{\partial y},
\quad    m_e\frac{dv_z}{dt}=qE_z-qv_yB_x,
\end{equation}
where $\Phi$ is the in-plane electrostatic potential of the ambipolar electric field $E_y(y,t)$ generated by plasma. In this model, the electrons move in the $y-z$ plane under the influence of both the electrostatic and inductive fields. They do not undergo pure cyclotron motion with only the RF magnetic field $B_x(y,t)$. Since the canonical momentum $m_ev_z+qA_z=const$ is seen to be conserved, the above equations reduce to 
\begin{equation}\label{yposition}
    \frac{d^2 y}{dt^2}=\frac{q}{m_e}\frac{\partial A_z}{\partial y}\left(v_{z0}-\frac{q}{m_e}(A_z-A_{z0})\right)-\frac{q}{m_e}\frac{\partial\Phi}{\partial y} \equiv-\frac{q}{m_e}\frac{\partial U}{\partial y},
\end{equation}
where $A_{z0}$ is the vector potential at the initial time and position of the electron being considered and the effective potential, $U(y,t)$, has been introduced as
\begin{equation}\label{effectivepotential}
    U=q\frac{A^2_z}{2}-(qA_{z0}+m_ev_{z0})A_z+\Phi.
\end{equation}
Figure~\ref{fig_effectivepotential} shows the time evolution of the effective potential and two representations of a trajectory obtained by numerically solving Eq.~\eqref{yposition}. The electrostatic potential and the vector potential waveforms entering Eq.~\eqref{yposition} are taken from the actual PIC simulation results. It is evident that the basic properties of the trajectory shown in Fig.~\ref{fig2_particletrace} (a) are reproduced by the 1D trajectory shown in Fig.~\ref{fig_effectivepotential} (b): in both cases the electrons originating near the coil are trapped in the effective potential well, undergoing a cyclotron motion in the $y-z$ plane, while drifting in $z$ direction. The maximum drift velocity calculated in the 1D model is around $4.8\times 10^6~ \text{m/s}$ (see Fig.~\ref{fig_effectivepotential} (b)), nearly the same as the maximum drift velocity in Fig.~\ref{fig2_particletrace} (a3). The bounce frequency $\Omega=\sqrt{qU''/m_e}$ in the effective potential well is nearly the same as the local electron cyclotron frequency $\Omega_B=qB/m_e$ (see discussion in Ref.~\cite{Sun_ICP_2025}). As we can see in Fig.~\ref{fig_effectivepotential} (a) and (b), the turning points of the bounce motion stay almost unchanged over time, so the average electron flux in $y$ direction vanishes in the lowest order. As a result, because the electron essentially take a bounce motion in the effective potential, \textit{the real adiabatic invariant of the trapped electron is the adiabatic invariant associated with the bounce motion $I=\int v_y dy$ instead of the magnetic moment $\mu=m_ev_{\perp}^2/2B$}. Therefore, because the turning points of the bounce motion stay almost unchanged (see Fig. \ref{fig_effectivepotential} (a)), the amplitude of $v_y$ oscillations stays almost unchanged as well. Remarkably, the phase portrait is not that different from a simple cyclotron motion in $v_y-v_z$ phase space (see Fig.~\ref{fig_effectivepotential} (c)). This justifies the simplification of using the cyclotron motion approximation applied in deriving the analytical expression for the electron current. }

\textcolor{black}{Now we move to consider the trapping condition for the electrons in the effective potential well. The trapping condition is
\begin{equation}\label{Eqreal_trappingcond}
    E=\frac{1}{2}m_ev_y^2+qU\leq min[qU(y=70mm),0],
\end{equation}
which means that electrons neither become lost to the wall nor escape to the bulk plasma. Since the thermal velocity is only about $0.15 qA_{max}/m_e$, most electrons are deeply trapped during most of the RF period. However, starting from $t=29.5\mu s$ (see Fig. \ref{fig_effectivepotential} (a)), the potential well no longer forms and the electrons start being lost to the bulk plasma. It is these electrons which eventually form the jet-like current identified in the simulations. To obtain a simpler analytical expression for the de-trapping condition, we simplify Eq.~\eqref{Eqreal_trappingcond} by requiring only the left turning point to exist (meaning that electrons do not go to the bulk plasma),
\begin{equation}
    \frac{1}{2}m_ev_{y0}^2+\frac{1}{2}q^2A^2_z-(qA_{z0}+m_ev_{z0})qA_z+q\Phi<0.
\end{equation}
Then, approximating the electrostatic potential as $q\Phi=q^2A_z^2/4m_e$ (justified in Fig. 35 of our accompanied paper \cite{Sun_ICP_2025}), we have approximately
\begin{equation}
    \left(\frac{\delta}{\rho_e}\right)^2+\frac{3}{2}\frac{\delta}{\rho_e}-1>0,
\end{equation}
where $\rho_e=m_ev_{th}/|qB_x|$ and $v_{y0}=v_{z0}=v_{th}$ is assumed, which becomes
\begin{equation}\label{trappinganaly}
    v_{th}\approx v_{z0}<\frac{|qA_z|}{m_e}\frac{1}{2}\approx \frac{|qB_x|\delta}{2m_e},
\end{equation}
which is exactly Eq. \eqref{estimte} in the main text.}

\textcolor{black}{In summary, the electrons are not only affected by the RF magnetic field, but also by the electrostatic potential in $x-y$ plane. The combination of them forms an effective potential that can trap the electrons. The adiabatic invariant is no longer magnetic momentum $\mu$, indicating that it is an essentially new physical process. This is why we always say ``trap'' and ``detrap'' in the main text.}

\clearpage
\bibliographystyle{apsrev4-2} 
\bibliography{apssamp}
\clearpage
\appendix

\end{document}